\RequirePackage{fix-cm}
\documentclass[smallextended]{svjour3}       
\smartqed  
\usepackage{graphicx}
\usepackage{amssymb}
\usepackage[mathscr]{eucal}
\usepackage{graphicx}

\usepackage{amsmath}
\usepackage{mathtools}

\usepackage[rightcaption]{sidecap}

\usepackage{graphicx} 
\graphicspath{ {images/} }

\usepackage{mathptmx}      
\usepackage{latexsym}
\usepackage{booktabs}
\usepackage{multirow}
\usepackage{tabularx}
\usepackage{caption}
\usepackage{latexsym}

 \usepackage{mathptmx}      
\usepackage{latexsym}
\usepackage{booktabs}
\usepackage{multirow}
\usepackage{tabularx}
\usepackage{caption}
\begin{document}

\title{Nonlinear oscillations of a point charge in the electric field of charged ring using a particular He's frequency-amplitude formulation
}

\titlerunning{Nonlinear oscillations of a point charge}        

\author{O. Gonz\'alez-Gaxiola      \and
        G. Chac\'on-Acosta \and J. A. Santiago
}


\institute{O. Gonz\'alez-Gaxiola \at
	Departamento de Matem\'aticas Aplicadas y Sistemas, Universidad Aut\'onoma Metropolitana-Cuajimalpa. Vasco de Quiroga 4871, Santa Fe, Cuajimalpa, 05300, Mexico D.F., Mexico\\
	\email{ogonzalez@correo.cua.uam.mx}           
	\and
	G. Chac\'on-Acosta \at
	Departamento de Matem\'aticas Aplicadas y Sistemas, Universidad Aut\'onoma Metropolitana-Cuajimalpa. Vasco de Quiroga 4871, Santa Fe, Cuajimalpa, 05300, Mexico D.F., Mexico
	\and
	J. A. Santiago \at
	Departamento de Matem\'aticas Aplicadas y Sistemas, Universidad Aut\'onoma Metropolitana-Cuajimalpa. Vasco de Quiroga 4871, Santa Fe, Cuajimalpa, 05300, Mexico D.F., Mexico
}

\date{Received: date / Accepted: date}

\maketitle

\begin{abstract}
In this paper, He's frequency-amplitude formulation with some choice of location points that improve accuracy is applied to determine the periodic solution for the nonlinear oscillations of a punctual charge in the electric field of charged ring. The results of the present study are valid for small and  large amplitudes of oscillation. The present method can be applied directly to highly nonlinear problems without any discretization,
linearization or restrictive assumptions. Finally, compared with the exact solution shows that the result obtained is of high accuracy and  we will do a comparative study with previous research done on the same problem, we will see that our approach is much better especially for large amplitude values.

\end{abstract}

\section{Introduction}
\label{s:Intro}
\noindent Nonlinear vibration arises everywhere in science,  engineering and other disciplines, since most phenomena in our world today, are essentially nonlinear and are described by nonlinear equations. It is very important in applications to have a version of the frequency (or period) to have a better understanding of the phenomena modeled through differential equations that contain terms with high nonlinearities, and a simple mathematical method is very useful for practical applications.\\
\noindent Recently, many different methods have been developed to obtain the approximate solutions of   nonlinear systems including the harmonic balance method \cite{Nay-1,Mic,Bel1,Bel-0}, the energy balance method \cite{Yil,Khan-1}, the Hamiltonian approach \cite{Yil-1,He-x0},  the use of special functions \cite{Zu-1,Zu-2}, the max-min approach \cite{He-x,Ze},  the variational iteration method \cite{Gan,He-1,He-2,He-3,Waz-1} and homotopy perturbation \cite{Bel-1,Bel-2,Gan-1,Gan-2,He-4,He-5,He-6}.  An excellent study, in which many of these techniques can be found in detail to solve nonlinear problems of oscillatory type can be seen in \cite{He-8}.\\
\noindent In dimensionless form, the equation of motion of a punctual charge  placed at a point on the ring axis, is given by the following nonlinear differential equation \cite{Bel-0,Bel-00}
\begin{equation}\label{Eq-1}
	\frac{d^{2}u}{dt^{2}}+\frac{u}{(1+u^{2})^{3/2}}=0,
\end{equation}
with the initial conditions
\begin{equation}\label{Ic}
	u(0)=A,\;\; u'(0)=0.
\end{equation}
Where $u$ is de position of the probe charge in units of the radius of the ring, $t$ is the time measured in terms of the natural frequency of the system.\\
The main purpose of the present work is to show how to apply the
frequency-amplitude formulation with a particular choice of location points, developed recently by Ji-Huan He based on an ancient Chinese algorithm in \cite{He-x1,He-x2,He-9,He-8},  and after improved by himself and other authors in \cite{He-10,He-11,Cai,Ren}, to obtain the approximate periodic solution of Eq.(\ref{Eq-1}).

\section{He's frequency-amplitude formulation}
\label{sn:AM}
\noindent In order to utilize He's amplitude frequency formulation, we choose two trial functions $u_{1}(t)=A\cos(t)$, $u_{2}(t)=A\cos(\omega t)$, which are respectively, the solutions of the following linear differential equations: 
\begin{align*} 
	u''+u &=  0 \\ 
	u''+\omega^{2} u &= 0,
\end{align*}
where $\omega$ is assumed to be the frequency of the nonlinear oscillator equation (\ref{Eq-1}). Substituting $u_{1}(t)$ and $u_{2}(t)$ into equation  (\ref{Eq-1}), we obtain, respectively, the following residuals
\begin{align*} 
	R_{1}(t) &=  \frac{A\cos(t)}{(1+A^{2}\cos^{2}(t))^{3/2}}-A\cos(t), \\ 
	R_{2}(t) &=  \frac{A\cos(\omega t)}{(1+A^{2}\cos^{2}(\omega t))^{3/2}}-A^{2}\omega^{2}\cos(\omega t).
\end{align*}
According to the method improvements made in \cite{Cai}, the approximate frequency can be obtained as follows:
\begin{equation}\label{Eq-2}
	\omega^{2}=\frac{R_{2}(t_2)-\omega^{2}R_{1}(t_1)}{R_{2}(t_2)-R_{1}(t_1)},
\end{equation}
where $t_1$ and $t_1$ are location points. The phase of the residuals is determined by the location points.\\
Recently, in \cite{Ren} an improvement in the manner of choosing these points was proposed. Such improvement works very well in the solution of certain problems, this is not the case of the problem studied here, as we will see in the conclusions and discussions  section.
In \cite{He-8,He-x1} the phase of the residuals was chosen as $0$ for simplicity, {\it i.e.} $t_{1}=0$ and $t_{2}=0$.
The corresponding residuals are as follows:
\begin{align} \label{Eq-3}
	R_{1}(0) &=  \frac{A}{(1+A^{2})^{3/2}}-A, \\ 
	R_{2}(0) &=  \frac{A}{(1+A^{2})^{3/2}}-A^{2}\omega^{2}.\label{Eq-4}
\end{align}
According to Eqs. (\ref{Eq-3}) and (\ref{Eq-4}), the frequency-amplitude formulation is

\begin{equation} \label{Eq-5}
	\begin{split}
		\omega^{2}(A) & = \frac{R_{2}(0)-\omega^{2}R_{1}(0)}{R_{2}(0)-R_{1}(0)} \\
		& = \frac{\Big(\frac{A}{(1+A^{2})^{3/2}}-A^{2}\omega^{2}\Big)-\omega^{2}\Big(\frac{A}{(1+A^{2})^{3/2}}-A\Big)}{\Big(\frac{A}{(1+A^{2})^{3/2}}-A^{2}\omega^{2}\Big)-\Big(\frac{A}{(1+A^{2})^{3/2}}-A\Big)}\\
		& =\frac{A(1-\omega^2)}{A(1-\omega^2)(1+A^{2})^{3/2}}=\frac{1}{(1+A^{2})^{3/2}}.
	\end{split}
\end{equation}
Therefore, the analytical approximate frequency $\omega$ as a function of $A$ is
\begin{equation}\label{Eq-6}
	\omega_{app}(A)=\frac{1}{\sqrt[4]{(1+A^{2})^{3}}}.
\end{equation}
From equation (\ref{Eq-6}) we obtain the following approximate periodic solution to (\ref{Eq-1})
\begin{equation}\label{Eq-8}
	u_{app}(t)=A\cos \left(\frac{1}{\sqrt[4]{(1+A^{2})^{3}}} t\right).
\end{equation}

\section{Exact and approximate solutions}
\label{sn:EvsA}
\noindent The nonlinear oscillator described by Eq. (\ref{Eq-1}) is a conservative system. Integrating Eq. (\ref{Eq-1}) and using the initial conditions Eq. (\ref{Ic}), the exact angular frequency as a function of $A$ was obtained in \cite{Bel-0}
\begin{equation}\label{Eq-9}
	\omega_{ex}(A) = \dfrac{2\pi}{\displaystyle \int_{0}^{1}\frac{4Adu}{\sqrt{(1+A^{2}u^{2})^{-1/2}-(1+A^{2})^{-1/2}}}}.
\end{equation}

\noindent In order to compare the approximate frequencies obtained in the present work and the exact ones given by Eq.  (\ref{Eq-9}) we consider the two cases. First we consider small amplitudes and we will see that our approximation is very good, in fact we will see that $\omega_{app}(A)$ of Eq. (\ref{Eq-6}) tends to be $\omega_{ex}(A)$ when $A\to + 0$. On the other hand we take into account large amplitudes and show that the approximation obtained by this method given by Eq. (\ref{Eq-6}) tends to be exactly $\omega_{ex}(A)$ when $A\to +\infty$.


\subsection{Case I: $0<A<1$}
\label{Sub1}
\begin{figure}[h!]
	\begin{center}
		\includegraphics[width=80mm, height=50mm, scale=1.0]{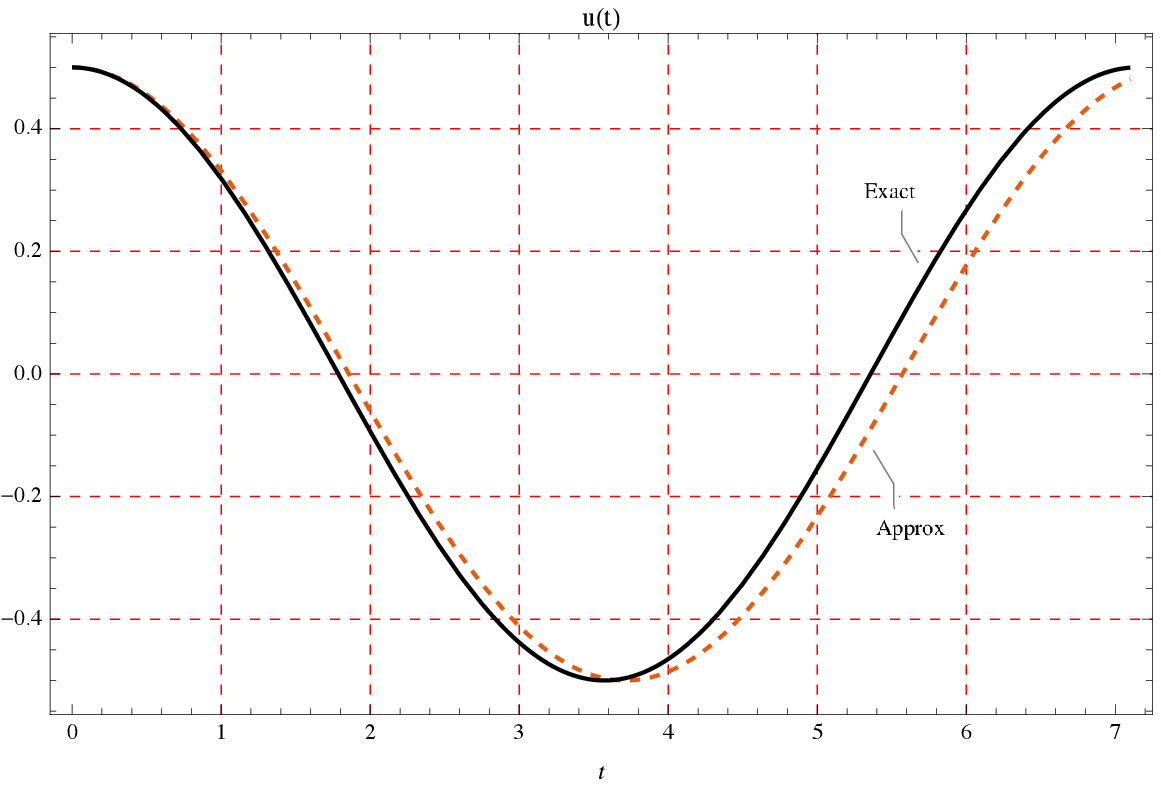}
	\end{center}
	\caption{Comparison of analytical approximation(dashed) and exact solution (black) for $A=1/2$\label{fi-0}}
\end{figure}

\noindent For small values of the amplitude A it is possible to tconsider the following approximation for $\omega_{ex}(A)$, as has been seen in \cite{Bel-0}
\begin{equation}\label{Eq-10}
	\omega_{ex}(A)\approx 1-\frac{9}{16}A^{2}+\frac{411}{1024}A^{4}-\frac{5147}{16384}A^{6}+\mathcal{O}(A^7).
\end{equation}
By taking into account our approximation made through He's frequency-amplitude formulation Eq. (\ref{Eq-6})  and $\omega_{ex}(A)$ from Eq. (\ref{Eq-10})  we can calculate  the Table \ref{tab-1} for small values of $A$.\\
\begin{table}[h!]
	\begin{center}
		\begin{tabular}{cccc}
			\cmidrule(r){1-4}
			$A$  & $\omega_{app}(A)$ Eq. (\ref{Eq-6}) &$\omega_{ex}(A)$ Eq. (\ref{Eq-10}) & $\frac{\omega_{app}}{\omega_{ex}}(A)$\\
			\midrule
			$1/1000$&$0.9999992500$&$0.9999994375$&$0.9999998124$\\
			$1/100$&$0.9999250066$&$0.9999437540$&$0.9999812515$\\
			$1/10$&$0.9925650290$&$0.9944148226$&$0.9981398169$\\
			$1/2$&$0.8458970108$&$0.8795518875$&$0.9617363373$\\
			\bottomrule
		\end{tabular}
	\end{center}
	\caption{Comparison between frequencies $\omega_{app}(A)$ and $\omega_{ex}(A)$ for small values of $A$. }
	\label{tab-1}   
\end{table}
\noindent In addition, considering that for small values of $A$ 
\begin{equation}\label{Eq-11}
	\omega_{app}(A)=\frac{1}{\sqrt[4]{(1+A^{2})^{3}}}\approx 1-\frac{3}{4}A^{2}+\frac{21}{32}A^{4}-\frac{77}{128}A^{6}+\mathcal{O}(A^7).
\end{equation}
Hence the ratio between $\omega_{ex}(A)$ and $\omega_{app}(A)$ can also be calculated as the following series
\begin{equation}\label{Eq-11-1}
	\frac{\omega_{app}(A)}{\omega_{ex}(A)} \simeq 1 - \frac{3}{16}A^{2} + \frac{153}{1024}A^{4} - \frac{17613}{16384}A^{6} + \mathcal{O}(A^7).
\end{equation}
In the exact limit  we have that
\begin{equation}\label{Eq-12}
	\lim\limits_{A\to 0^{+}}\frac{\omega_{app}(A)}{\omega_{ex}(A)}=1,
\end{equation}
and hence, $\omega_{app}(A)\to \omega_{ex}(A)$ as $A\to 0^{+}$.\\
In Fig. \ref{fi-0} we plot the comparison between exact solution $u_{ex}=A\cos(\omega_{ex}(A))$ for $A=1/2$. Even though around $A = 1/2$ the error is large, as one can see in Table \ref{tab-1}, the approximation is acceptable.

\subsection{Case II: $A\geq  1$}
\label{Sub2}
\begin{figure}[h!]
	\begin{center}
		\includegraphics[width=80mm, height=50mm, scale=1.0]{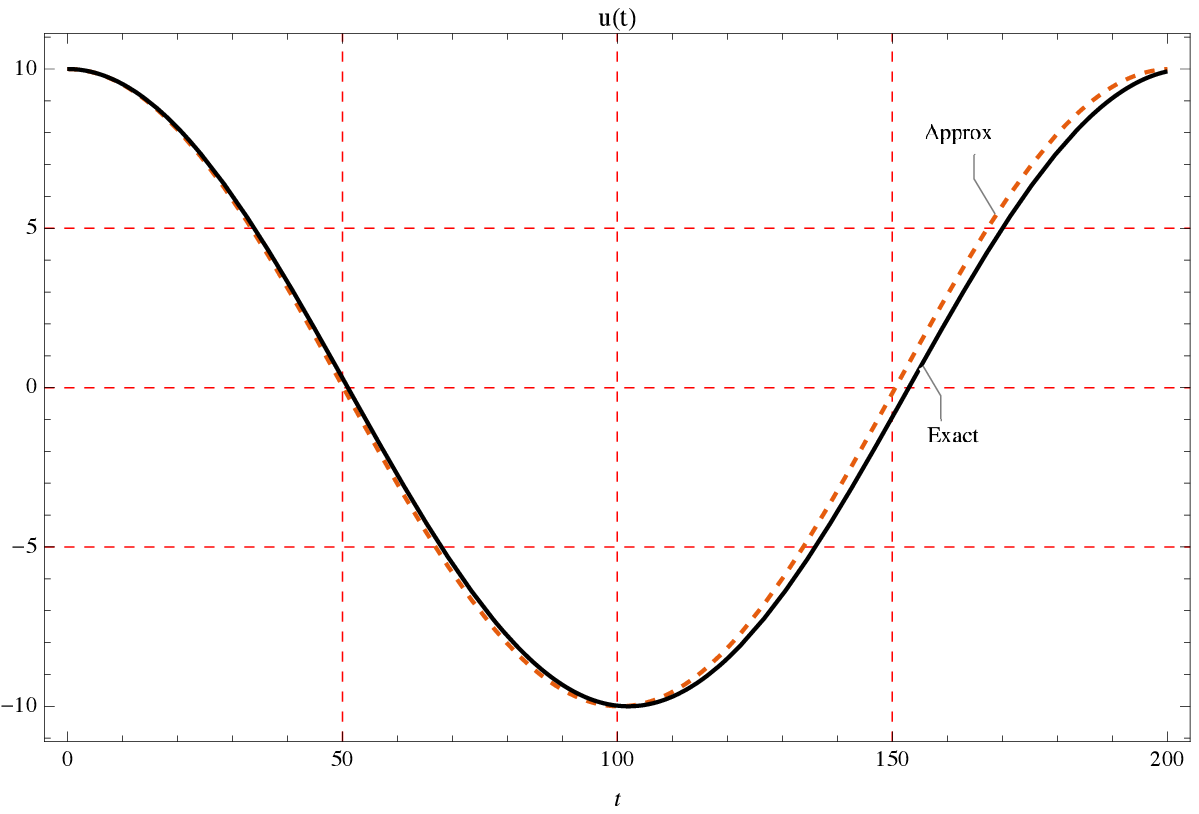}
	\end{center}
	\caption{Comparasion of analytical approximation(dashed) and exact solution (black) for $A=10$\label{fi-00}}
\end{figure}

\noindent For large amplitudes we calculate the following  values with the approximation made through He's frequency-amplitude formulation Eq. (\ref{Eq-6})  and the exact result $\omega_{ex}(A)$ given by Eq. (\ref{Eq-9})
\begin{table}[h!]
	\begin{center}
		\begin{tabular}{cccc}
			\cmidrule(r){1-4}
			$A$  & $\omega_{app}(A)$ Eq. (\ref{Eq-6}) &$\omega_{ex}(A)$ Eq. (\ref{Eq-9}) & $\frac{\omega_{app}}{\omega_{ex}}(A)$\\
			\midrule
			$1$&$0.5946035575$&$0.4704457120$&$1.2639153516$\\
			$5$&$0.0868500332$&$0.0831210434$&$1.0448621628$\\
			$10$&$0.0313876621$&$0.0308692709$&$1.0167931144$\\
			$50$&$0.0028275788$&$0.0028231732$&$1.0015605432$\\
			$100$&$0.0009999250$&$0.0009993712$&$1.0005541029$\\
			$1000$&$0.0000316227$&$0.0000316221$&$1.0000175879$\\
			\bottomrule
		\end{tabular}
	\end{center}
	\caption{Comparison between frequencies $\omega_{app}(A)$ and $\omega_{ex}(A)$ for large values of $A$ }
	\label{tab-2}   
\end{table}
\noindent In this case, we calculate again the ratio between both frequencies approximated by exact ones, as in \cite{DQ}. In our calculations we need to take into account that the following integral do converge
\begin{equation}\label{Eq-13}
	\int_{0}^{1}\frac{4\sqrt{u}du}{\sqrt{1-u}}=2\pi.
\end{equation}
In the asymptotic limit for the ratio is calculated as follows
\begin{equation} \label{Eq-14}
	\begin{split}
		\lim\limits_{A\to +\infty}\frac{\omega_{app}(A)}{\omega_{ex}(A)} & = \lim\limits_{A\to +\infty}\frac{\frac{1}{\sqrt[4]{(1+A^{2})^{3}}}}{\dfrac{2\pi}{\Big(\displaystyle \int_{0}^{1}\frac{4Adu}{\sqrt{(1+A^{2}u^{2})^{-1/2}-(1+A^{2})^{-1/2}}}\Big)}}\\
		& = \lim\limits_{A\to +\infty}{\dfrac{\displaystyle \int_{0}^{1}\frac{4Adu}{\sqrt{(1+A^{2}u^{2})^{-1/2}-(1+A^{2})^{-1/2}}}}{2\pi\sqrt[4]{(1+A^{2})^{3}}}}\\ &
		=\frac{1}{2\pi} \int_{0}^{1}\frac{4\sqrt{u}du}{\sqrt{1-u}}=1.
	\end{split}
\end{equation}
Therefore, $\omega_{app}(A)\to \omega_{ex}(A)$ as $A\to +\infty$.\\
In Fig. \ref{fi-00} we compare the exact solution $u_{ex}=A\cos(\omega_{ex}(A))$ for $A=10$ and the corresponding approximation. We can appreciate that the semianalytical solution is quite acceptable.\\

\vspace{0.1in}

\noindent From Tables \ref{tab-1}, \ref{tab-2}  and Eqs. (\ref{Eq-12}), (\ref{Eq-14}), it can be observed that Eq. (\ref{Eq-6}) yield excellent analytical approximate frequencies for both, very small and very large values of oscillation amplitude $A$.

\section{Conclusions and Discussions}
\label{Con}
\begin{figure}[h!]
	\begin{center}
		\includegraphics[width=85mm, height=58mm, scale=1.0]{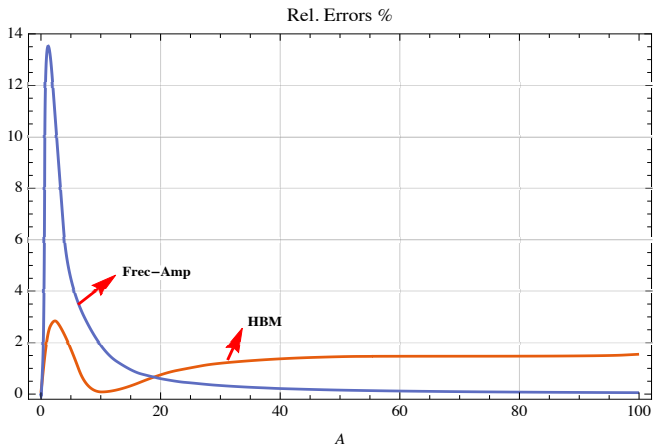}
	\end{center}
	\caption{Relative errors with the method (Amp-Frec) used in the present study (blue) and the HMB (orange) in \cite{Bel-0}\label{fi1}}
\end{figure}

\noindent In this section we study the precision and validity of the solution method employed in the present work. We compare our results with the exact solution obtained in \cite{Bel-0}, all numerical calculations have been made with the help of the software MATHEMATICA. A comparative analysis of the error given by the present method and the error of the best approximation obtained by the Harmonic balancing approach \cite{Bel-0}, is made.\\
The exact solution for the frequency $\omega_{ex}(A)$ is given by Eq. (\ref{Eq-9}) and  we can obtain an asymptotic expansion for large amplitudes as was done in \cite{Bel-0}, which is given by 
\begin{equation}\label{Eq-15}
	\omega_{ex}(A)\approx \frac{\sqrt{2}}{A^{3/2}}.
\end{equation}
There have been obtained a good approximation for $\omega(A)$ in \cite{Yil} through the Energy Balance Method (EBM), which conclusion is 
\begin{equation}\label{Eq-16}
	\omega_{EBM}(A)= \frac{2}{A^{3/2}}\sqrt{\sqrt{2}-1}
\end{equation}
where, considering Eq. (\ref{Eq-15}) we can obtain the large amplitude limit of the ratio of frequencies 
\begin{equation}\label{Eq-17}
	\lim\limits_{A\to \infty}\frac{\omega_{HBM}(A)}{\omega_{ex}(A)}=0.9101797.
\end{equation}
From Eq. (\ref{Eq-17}) one can see that the relative percentage error tends to $8.9\%$ for $A\to\infty$.\\
In addition, the problem was studied in \cite{Vali} with the same method. The main difference with the present work is that in \cite{Vali} the points were chosen following Reng-Gui \cite{Ren} and Geng-Cai \cite{Cai}, in such work the best approximation obtained was the following
\begin{equation}\label{Eq-18}
	\omega_{R-G}(A)= \frac{1.3337481}{A^{3/2}}
\end{equation}
where, by considering Eq. (\ref{Eq-15}) they found
\begin{equation}\label{Eq-19}
	\lim\limits_{A\to \infty}\frac{\omega_{R-G}(A)}{\omega_{ex}(A)}=0.9431023.
\end{equation}
From Eq. (\ref{Eq-19}) it is observed that the relative percentage error tends to $5.6\%$ in the limit $A\to\infty$.\\
The best known approximation was obtained by the Harmonic Balance Method (HBM) in \cite{Bel-0}, whose approximation is 
\begin{equation}\label{Eq-20}
	\omega_{HBM}(A)= \frac{1.39197}{A^{3/2}}
\end{equation}
from where
\begin{equation}\label{Eq-21}
	\lim\limits_{A\to \infty}\frac{\omega_{HBM}(A)}{\omega_{ex}(A)}=0.984273.
\end{equation}
The above, is the best known approximation so far, with which we have a percentage relative error of only $1.57\%$  in the limit $A\to\infty$.\\
By considering the data of tables \ref{tab-1} and \ref{tab-2}, we calculate the percentage relative error for small and large amplitudes $A$, the corresponding values are shown in Table \ref{tab-3}.
%
\begin{table}[h!]
	\begin{center}
		\begin{tabular}{cccc}
			\cmidrule(r){1-4}
			$A$  & Relative errors \% &A &  Relative errors \%\\
			\midrule
			$1/1000$&$0.0000181$&$10$&$1.6793114$\\
			$1/100$&$0.0018742$&$20$&$0.6085781$\\
			$1/10$&$0.1860181$&$50$&$0.1560544$\\
			$1/2$&$3.8263660$&$100$&$0.0554103$\\
			$1$&$13.318421$&$1000$&$0.0017588$\\
			\bottomrule
		\end{tabular}
	\end{center}
	\caption{Relative errors \% obtained in the present study  for some small and large values of $A$ }
	\label{tab-3}   
\end{table}
\noindent In the present work, from Eq. (\ref{Eq-14}) we obtain the limit  $\lim\limits_{A\to +\infty}\frac{\omega_{app}(A)}{\omega_{ex}(A)}=1$, From where the percentage relative error tends to zero when $A\to\infty$ as we can see in the figure \ref{fi1}, which improves the best known approximation  $\omega_{HBM}(A)$. Also, in the figure \ref{fi1}  it can be seen that for $0<A<1$  the harmonic balance method used in \cite{Bel-0} is a better approximation for the solution of the present problem.\\
\noindent Additionally, we must clarify that the center of the ring presents instability and therefore the oscillations of the charge around this point are not possible. To solve this problem it could be possible to consider
that the punctual charge is on a finite conducting wire placed
along the axis of the ring as it is done as in \cite{Luc}, this is also considered in \cite{Bel-00}.

\noindent Finally, it is concluded from the previous discussion that the alternative approach presented in this paper for the selection of location points in He's frecuency-amplitude formulation, gives a high precision solution for the nonlinear oscillations of a ponit charge in the electric field of charged ring, specially for large values of $A\gg1$ or very small $ A \ll 1 $.



\end{document}